\documentclass{article}
\usepackage{spconf,amsmath,graphicx}
\usepackage{amsthm,amsfonts,bm,xspace}
\usepackage{amsfonts}
\usepackage{algorithmic}
\usepackage[algo2e]{algorithm2e}

\title{cDPMSR: Conditional Diffusion Probabilistic Models for \\
Single Image Super-Resolution }
\name{Axi Niu$^{1}$, Kang Zhang$^2$, Trung X. Pham$^2$, Jinqiu Sun$^1$*, Yu Zhu$^1$, In So Kweon$^2$, Yanning Zhang$^1$
\thanks{This work was funded in part by the Project of the National Natural Science Foundation of China under Grant 61871328, Natural Science Basic Research Program of Shaanxi under Grant 2021JCW-03, as well as the Joint Funds of the National Natural Science Foundation of China under Grant U19B2037.). (*Corresponding author: Jinqiu Sun.)}}

\address{$^1$Northwestern Polytechnical University \\$^2$ Korea Advanced Institute of Science and Technology (KAIST)
}

\usepackage{graphicx}
\usepackage{amsmath}
\usepackage{stfloats}
\usepackage{multirow}
\usepackage{xcolor}
\usepackage{booktabs}
\usepackage{CJKutf8}
\usepackage{pifont}
\usepackage{mathrsfs}
\usepackage{makecell}
\usepackage{amsmath}
\usepackage{algorithm}
\usepackage{color}

\newcommand{\eg}{\emph{e.g.}}
\newcommand{\ie}{\emph{i.e.}}

\begin{document}
\maketitle

\begin{abstract}
Diffusion probabilistic models (DPM) have been widely adopted in image-to-image translation to generate high-quality images. Prior attempts at applying the DPM to image super-resolution (SR) have shown that iteratively refining a pure Gaussian noise with a conditional image using a U-Net trained on denoising at various-level noises can help obtain a satisfied high-resolution image for the low-resolution one. To further improve the performance and simplify current DPM-based super-resolution methods, 
we propose a simple but non-trivial DPM-based super-resolution post-process framework,\ie, cDPMSR. 
After applying a pre-trained SR model on the to-be-test LR image to provide the conditional input, we adapt the standard DPM to conduct conditional image generation and perform super-resolution through a deterministic iterative denoising process.
Our method surpasses prior attempts on both qualitative and quantitative results and can generate more  photo-realistic counterparts for the low-resolution images with various benchmark datasets including Set5, Set14, Urban100, BSD100, and Manga109. \textit{Code will be published after accepted}.
\end{abstract}

\vspace{-10pt}
\section{Introduction}
\label{sec:intro}
\vspace{-10pt}

Over the years, single image super-resolution (SISR) has drawn active attention due to its wide applications in computer vision, such as object recognition, remote sensing, and so on.
SISR aims to obtain a high-resolution (HR) image containing great details and textures from a low-resolution (LR) image by an SR method, which is a classic ill-posed inverse problem~\cite{niu2022ms2net}. To establish the mapping between HR and LR images, various CNN-based methods had been proposed. Among them, methods based on the deep generative model have become one of the mainstream, mainly including GAN-based~\cite{wang2018esrgan,soh2019natural,tian2022generative} and flow-based methods~\cite{lugmayr2020srflow,liang2021hierarchical,wolf2021deflow}, which have shown convincing image generation ability.

GAN-based SISR methods~\cite{wang2018esrgan,soh2019natural,tian2022generative} used a generator and a discriminator in an adversarial way to encourage the generator to generate realistic images. Specifically, the generator generates an SR result for the input, and the discriminator is used to distinguish if the generated SR is true. It combines content losses (\eg, $L_{1}$ or $L_{2}$) and adversarial losses to optimize the whole training process. Due to their strong learning abilities, GAN-based methods become popular for image SR tasks~\cite{tian2022generative,bell2019blind,emad2021dualsr}. However, these methods are easy to meet mode collapse and the training process is hard to converge with complex optimization~\cite{metz2016unrolled,ravuri2019classification} and adversarial losses often introduce artifacts in SR results, leading to large distortion~\cite{lugmayr2021ntire,whang2022deblurring}. Another line of methods based on deep generative models is flow-based methods, which directly account for the ill-posed problem with an invertible encoder~\cite{kim2021noise,liang2021flow,saharia2022image,laroche2022bridging}. It transforms a Gaussian distribution into an HR image space instead of modeling one single output and inherently resolves the pathology of the original ”one-to-many” SR problem. Optimized by a negative loglikelihood loss, these methods avoid training instability but suffer from extremely large footprints and high training costs due to the strong architectural constraints to keep the bijection between latents and data~\cite{saharia2022image}. 

\begin{figure}[!t]
\centering	
    \includegraphics [width=0.92\linewidth]{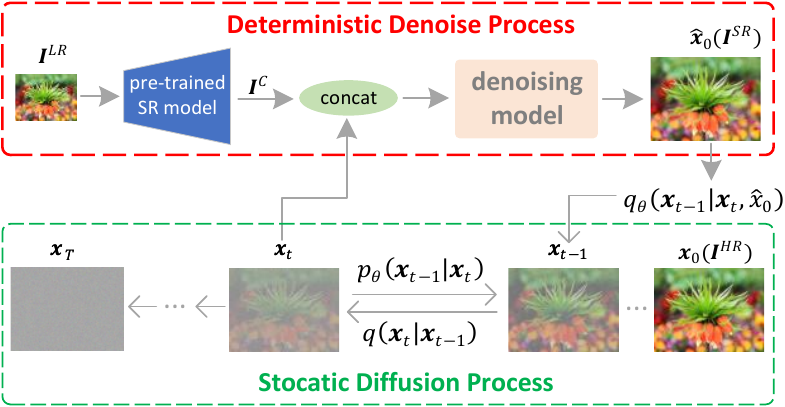}
    \vspace{-8pt}
    \caption{Illustration of our method. The model contains a stochastic forward diffusion process which gradually adds noise to an $\bm I^{HR}$ image. And a deterministic denoise process is applied to recover high-resolution and realistic images $\bm I^{SR}$  corresponding to $\bm I^{LR}$  images.}
    \label{fig:overview}
    \vspace{-10pt}
\end{figure}

Lately, the adoptions of diffusion probabilistic models (DPM) have shown promising results in image generative tasks~\cite{ho2020denoising}.
Furthermore, the prior attempts~\cite{li2022srdiff,saharia2022image} at applying the DPM to image SR have also proved their effectiveness for obtaining satisfied SR images.
In~\cite{li2022srdiff}, the authors propose a two-stage SR framework. First, they design an SR structure and pre-train it to obtain a conditional input for the DPM process. Then they redesign the U-net in DPM. The training process of this method is relatively complicated, and it does not consider combining existing pre-trained SR models, such as EDSR~\cite{lim2017enhanced}, RCAN~\cite{zhang2018image}, and SwinIR~\cite{liang2021swinir}. Similarly, ~\cite{saharia2022image} apply the bicubic up-sampled LR image as the conditional input directly. Different from them, our method leverage the development of the current SISR methods to provide more plausible conditional inputs. And we adapt a deterministic sampling way in the inference phase to make the restoration of the SR image better and faster.

Our work is most similar to~\cite{li2022srdiff} which is the first to apply DPM to the SR task. In this work, we propose a simple but non-trivial SR post-process framework for image SR based on the conditional diffusion model,\ie, cDPMSR. Unlike the existing techniques, our cDPMSR adopts the pre-trained SR methods to provide the conditional input, which is more plausible than the one used in~\cite{saharia2022image,li2022srdiff}. And it brings significant improvement in perceptual quality over existing SOTA methods across multiple standard benchmarks. By simply concatenating a Gaussian noise and the conditional input and using MAE loss (\ie, $L_{1}$ ) to optimize the diffusion model, our method makes the training process more concise compared with ~\cite{saharia2022image,li2022srdiff}.
Our contributions are summarized as follows: 
\begin{itemize}
    \vspace{-4pt}
    \item To the best of our knowledge, we are the first to propose an SR post-process framework based on the existing SR models and diffusion probabilistic model.
    \vspace{-4pt}
    \item Compared to existing SOTA SR methods, our cDPMSR achieves superior perceptive results and can generate more photo-realistic SR results.
    \vspace{-4pt}
    \item Compared with existing DPM-based SR methods, our cDPMSR adopts a deterministic sampling way in the inference phase, which helps to obtain a better balance between distortion and perceptual quality.
\end{itemize}

\vspace{-14pt}
\section{Proposed Method}
\vspace{-6pt}
\label{sec:pagestyle}

Fig.\ref{fig:overview} illustrates the whole process of our cDPMSR. The following section introduces the method in detail.

\vspace{-10pt}
\subsection{Stocatic Diffusion Process}
Given a SISR dataset $(\bm I^{HR}, \bm I^{LR})\sim D$, we adopt the diffusion probabilistic model (DPM)
~\cite{ho2020denoising,bansal2022cold} to map a normal Gaussian noise $\bm x_T \sim\mathcal{N}(0, \bm 1)$ to a high-resolution image $ I^{HR}$ with a corresponding conditional image $\bm I^C$. We will talk about the choice of the conditional image later. The DPM contains latent variables $\bm x=\{\bm x_t|t=0,1,...,T\}$, where $\bm x_0 = I^{HR}$, $ \bm x_T = \mathcal{N}(0, \bm 1)$. And it is defined on a noise schedule $\alpha_t$ and $\sigma_t$ such that $\lambda_t=\alpha_t^2/\sigma_t^2$, the signal-to-noise-ratio, is monotonically decreasing with $t$.

\noindent\textbf{Forward stochastic diffusion process.} We define the forward process $q(\bm x_t| \bm I^{HR}):=q(\bm x_t| \bm x_0)$ of DPM with a Gaussian process by the Markovian structure:
\vspace{-6pt}
\begin{equation}
\begin{aligned}
     q(\bm x_t| \bm x_0)&=\mathcal{N}(\bm x_0; \alpha_t\bm x_0,\sigma_t^2\bm1)\\
     q(\bm x_t|\bm x_{t-1}) &= \mathcal{N}(\bm x_t; (\alpha_t/\alpha_{t-1})\bm x_{t-1},(1-\frac{\lambda_t}{\lambda_{t-1}})\sigma_t^2\bm1).
\end{aligned}
\end{equation}

The forward process gradually adds noise into an image $\bm x_0$ to generate latent variables $\bm x_1,...,\bm x_T$ for the original image $\bm x_0$. With the Gaussian distribution reparameterization trick, we can write the latent variable $x_t$ as:

\begin{equation}
\label{eq:forwardxt}
    \bm x_t = \alpha_t \bm x_0 + \sigma_t \bm\epsilon, \epsilon\sim\mathcal{N}(0,\bm1).
\end{equation}
\vspace{-1pt}
Following~\cite{ho2020denoising} we set $\alpha_t=\cos{0.5\pi \frac{t}{T}}$ and $\sigma_t=\sqrt{1-\alpha_t^2}$.

\noindent\textbf{Model training.} The optimization target of DPM is denoising $\bm x_t\sim p(\bm x_t|\bm x_0)$ to get estimated $ \hat{\bm x}_0$ with a U-Net $f_\theta(x_t, t, I^C):=\hat{\bm x_0}\approx \bm x_0$. Same with~\cite{ho2020denoising}, we use the following loss function to train the model:
\begin{equation}
     L := \mathbb E_{t,(\bm x_0, I^C), \epsilon}[\| \bm x_0 - f_\theta(\alpha_t \bm x_0 + \sigma_t \bm\epsilon, t, I^C) \|],
\end{equation}
where $t$ is uniformly sampled between 1 and $T$. In~\cite{ho2020denoising,bansal2022cold} the above loss function is justified as optimizing the usual variational bound on negative log-likelihood with discarding the loss weighting. Here, different with~\cite{ho2020denoising} we add an additional input $I^C$ as the conditional image to guide the model to keep the same content with $I^C$ during the denoising process. And different from current DPM-based super-resolution methods~\cite{saharia2022image,li2022srdiff} which reverses the diffusion process by estimating noise, we directly let the model predict the image.

\vspace{-8pt}
\begin{algorithm}[!htbp]
  \caption{cDPMSR}
  \label{alg:cDPMSR}
\small
  \SetKwInOut{Training}{Training}\SetKwInOut{Inference}{Inference}
  \SetKwData{Input}{Input:}\SetKwFunction{Uniform}{Uniform}
  \SetKwData{Require}{Require:}
\Training {train predictor $f_\theta$}
\Input Dataset $D$, schedule $\alpha_t,\sigma_t$, timesteps $T$, pretrained super resolution model $\phi_\theta$
  \BlankLine
  \begin{algorithmic}[1]
    \REPEAT
      \STATE $(\bm I^{HR}, \bm I^{LR}) \sim D$, $t \sim$ \Uniform{$\{1,...,T\}$}, \\${\bm \epsilon}\sim\mathcal{N}(\mathbf{0},\mathbf{I} )$
      \STATE $\bm I^{C} = \phi_\theta(I^{LR})$
      \STATE $\bm x_t = \alpha_t\bm I^{HR}+\sigma_t\bm \epsilon$
      \STATE Take a gradient descent step on\\
       $\qquad \nabla_\theta \|\bm I^{HR} - f_\theta(\bm x_t, t, \bm I^{C})  \|$ 
    \UNTIL{converged} 
  \end{algorithmic}
  \Inference { super resolve $\bm I^{LR}$}
\Input trained predictor $f_\theta$, pretrained super resolution model $\phi_\theta$
\BlankLine
    \begin{algorithmic}[1]
    \STATE $\bm I^{C} = \phi_\theta(I^{LR})$
    \STATE ${\bm x_T}\sim\mathcal{N}(\bm{0},\bm{1} )$
    \FOR{$t=T,...,1$}
    \STATE ${\bm\epsilon}\sim\mathcal{N}(\mathbf{0},\mathbf{I} )$ if $t>1$, else $\bm\epsilon=\mathbf{0}$
    \STATE $\bm x_{t-1} = \alpha_{t-1} f_\theta(x_t, t, I^C) + \sigma_{t-1} (\bm x_t -\alpha_t\hat{\bm x_0} )/\sigma_t$
    \ENDFOR 
  \end{algorithmic}
\end{algorithm}

\vspace{-18pt}
\subsection{Deterministic Denoise Process}
\textbf{Trained model sampling.} Different from~\cite{li2022srdiff,saharia2022image} sampling via a stochastic way, we adopt a deterministic manner to conduct the reverse process $p_\theta(\bm x_{t-1}|\bm x_t)$, which is an implicit probabilistic model~\cite{song2020denoising}. Compared with the sampling strategy used in~\cite{li2022srdiff,saharia2022image}, it can achieve higher quality images with less inference time which is critical for this task. Given the image $\bm x_t$ at step $t$, we can write the generation process of $\bm x_{t-1}$ via a forward posterior $q_\theta(\bm x_{t-1}|\bm x_t,\bm \hat{x}_0)$ as follows:
\vspace{-6pt}
\begin{equation}
\begin{aligned}
p_\theta(\bm x_{t-1}|\bm x_t)&=q_\theta(\bm x_{t-1}|\bm x_t,\bm \hat{x}_0)\\
&= \mathcal{N}(\alpha_{t-1} \hat{\bm x}_\theta(\bm x_t, t) + \sigma_{t-1} \hat{\bm z}, \bm 0)\\
\bm x_{t-1} &= \alpha_{t-1} \hat{\bm x}_\theta(\bm x_t, t) + \sigma_{t-1} \hat{\bm z},
\end{aligned}
\end{equation}
where $\hat{\bm x}_0$ is predicted with $f_\theta(x_t, t, I^C)$ and $\hat{\bm z}$ is the estimated noise which can be calculated with Equation~\ref{eq:forwardxt}, $\hat{\bm z}=(\bm x_t -\alpha_t\hat{\bm x}_0)/\sigma_t$. Integrating the above equation we have:
\begin{equation}
    \bm x_{t-1} = \alpha_{t-1} f_\theta(x_t, t, I^C) + \sigma_{t-1} (\bm x_t -\alpha_t\hat{\bm x}_0 )/\sigma_t.
\end{equation}

\noindent\textbf{Conditional image choice.} To get realistic super-resolution images, \cite{li2022srdiff, saharia2022image} also introduced DPM with conditional denoising on a pre-trained feature extractor or a bicubic upsampled image on low-resolution image. In this work, we leverage the power of the current development of SISR to provide a better conditional image. Specifically, given a low-resolution image $I^{LR}$ and a pre-trained super-resolution model $\phi_\theta$, we generate our conditional image by $I^C=\phi_\theta(I^{LR})$, which has been proved to be more plausible for obtaining results with better perceptual quality in ablation study~\ref{ab:different_conditional_inputs}. Algorithm~\ref{alg:cDPMSR} shows the whole process of our cDPMSR.
\begin{figure*}[!ht]
\centering	
    \includegraphics [width=0.85\linewidth]{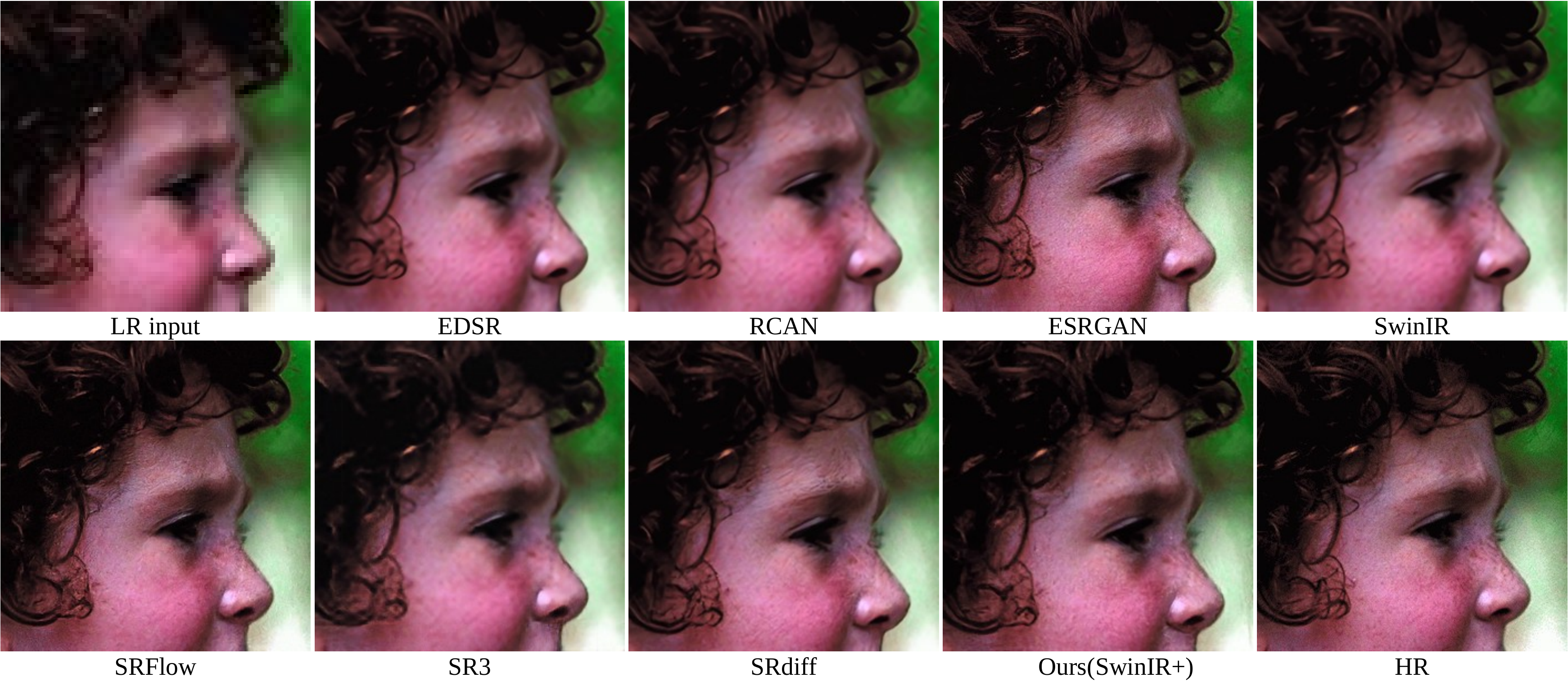}
    \vspace{-12pt}
    \caption{Qualitative comparison with SOTAs performed on `head' from Set5 ($\times$4 scale, best view in zoomed-in.)}
\vspace{-8pt}
    \label{fig:main_figure_1}
\end{figure*}
\vspace{-8pt}
\begin{figure*}[!ht]
\centering	
    \includegraphics [width=1\linewidth]{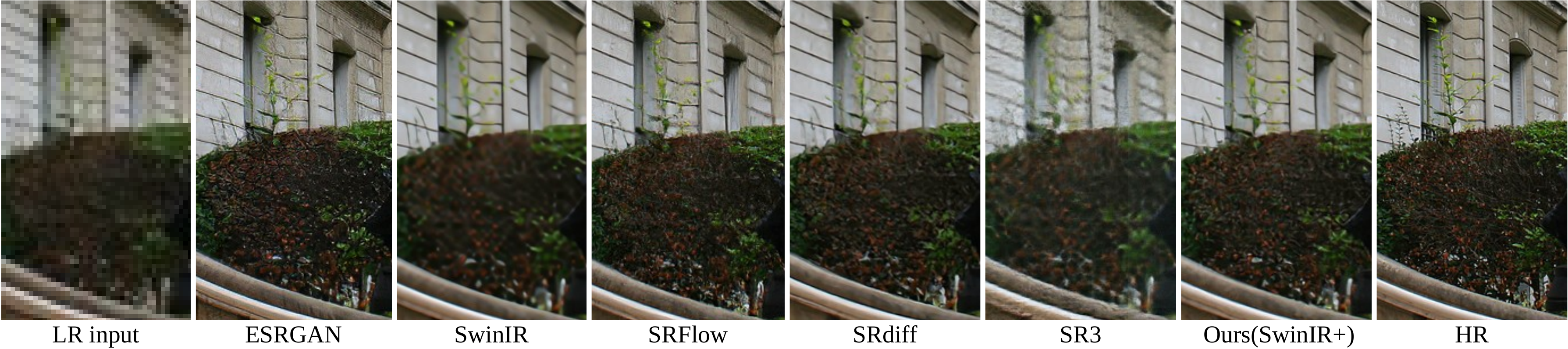}
    \vspace{-20pt}
    \caption{Qualitative comparison with SOTAs performed on `img003 ' from Urban100 ($\times$4 scale. Cropped and zoomed in for a better view.)}
    \label{fig:main_figure_2}
    \vspace{-18pt}
\end{figure*}

\section{Experiments}
\label{sec:typestyle}

\subsection{Experimental Settings}
\textbf{Implementation Details.} We use 800 image pairs in DIV2K as the training set. We take public benchmark datasets,~\ie, Set5, Set14, Urnban100, BSD100, and Manga109 as the test set to compare with other methods. For the diffusion model, we set $T=1000$ for training and $T=100$ during inference time. We take the pre-trained super-resolution models (SwinIR~\cite{liang2021swinir}, EDSR~\cite{lim2017enhanced}, and RCAN~\cite{zhang2018image}) to provide the initial super-resolution image,\ie the conditional input image. The conditional diffusion model is trained with Adam optimizer and batch size 16, learning rate $1\times 10^{-4}$ for 300k steps. The architecture of the model is the same as the one in~\cite{saharia2022image}. 

\noindent\textbf{A note on metrics.} The previous study has shown the distortion and perceptual quality are at odds with each other, and there is a trade-off between them~\cite{blau2018perception}. Since our work focuses on the perceptual quality, except the distortion metrics: PSNR and SSIM, we also provide perceptual metrics:  LPIPS~\cite{zhang2018unreasonable}
to show that our method can generate better perceptual results than other methods. LPIPS is recently introduced as a reference-based image quality evaluation metric, which computes the perceptual similarity between the ground truth and the SR image.
\vspace{-12pt}
\subsection{Quantitative and qualitative results}

To verify the effectiveness of our cDPMSR, we select some SOTA generative methods to conduct the comparative experiments, including ESRGAN~\cite{wang2018esrgan},  SRFlow~\cite{lugmayr2020srflow}, SRDiff~\cite{li2022srdiff}, SR3~\cite{saharia2022image}. We selected EDSR~\cite{lim2017enhanced}, RCAN~\cite{zhang2018image}, and SwinIR~\cite{liang2021swinir} to provide conditional input, respectively. Therefore, we report three cases for our cDPMASR,~\ie, \textbf{EDSR+}, \textbf{RCAN+}, and \textbf{SwinIR+}. All the results are obtained by the provided codes or from the publicized papers. As shown in Fig.\ref{fig:main_figure_1} and~\ref{fig:main_figure_2}, the results of EDSR, RCAN, and SwinIR are so smooth that some details are missed. Because these methods are PSNR-directed and they all focus on obtaining results with good  distortion~\cite{blau2018perception} 
and they certainly got good PSNR results in Tab.\ref{tab:results_quantitative}. 
It seems the results generated by ESRGAN in Fig.~\ref{fig:main_figure_1} and~\ref{fig:main_figure_2} include more details than other methods, but it introduces too many false artifacts compared to the ground truth. And the results of SRflow seem a little noisy. 
With better conditional input, our method exhibits superior performance on both quantitative and qualitative results than SR3~\cite{saharia2022image}. Though SRdiff obtains comparable numeric results in Tab.\ref{tab:results_quantitative}, the visual results of our cDPMSR are closer to ground truths (Especially, the forehead in Fig.\ref{fig:main_figure_1} and the plants in Fig.\ref{fig:main_figure_2}).

\begin{table*}[!ht]
\begin{center}
\caption{Results for 4$\times$SR on Set5, Set14, BSD100, Urban100, and Manga109. The best three perceptual results are highlighted in \textcolor{red}{red}, \textcolor{green}{green}, and \textcolor{blue}{blue} colors, respectively. The \textbf{bold} represents the best distortion result among generative-based methods. } 
\label{tab:results_quantitative}
\resizebox{0.9\hsize}{!}{
\begin{tabular}{cc|c|c|c|c|c|c|c|c|c|c}
\toprule
\hline
\multicolumn{2}{c|}{\multirow{2}{*}{Method} } 
& \multirow{2}{*}{EDSR}   & \multirow{2}{*}{RCAN}   & \multirow{2}{*}{ESRGAN} & \multirow{2}{*}{SwinIR} & \multirow{2}{*}{SRFlow} & \multirow{2}{*}{SRDiff} & \multirow{2}{*}{SR3} &  \multicolumn{3}{c}{\textbf{Ours}} \\ \cline{10-12}
&&&&&&&&& \textbf{EDSR+} & \textbf{RCAN+} & \textbf{SwinIR+}\\\midrule
\multicolumn{1}{c|}{\multirow{3}{*}{Set5}}     & LPIPS$\downarrow $ & 0.0922 & 0.1096 & \textcolor{green}{0.0596} & 0.0899 & 0.0767 & 0.0770 & 0.1265 & 0.0619 & \textcolor{blue}{0.0606} &\textcolor{red}{0.0564}  \\ \cline{2-12} 
\multicolumn{1}{c|}{}                          & PSNR$\uparrow $  & 32.43  & 32.64  & 30.46  & 32.69  & 28.35  & 30.94  & 27.31  & 30.78  & 30.84  &\textbf{31.03} \\ \cline{2-12} 
\multicolumn{1}{c|}{}                          & SSIM$\uparrow $  & 0.8985 & 0.9002 & 0.8516 & 0.9018 & 0.8138 &\textbf{0.8738} & 0.7844 & 0.8684 & 0.8645 & 0.8676  \\ \midrule
\multicolumn{1}{c|}{\multirow{3}{*}{Set14}}     & LPIPS$\downarrow $ & 0.1445 & 0.1387 & \color{blue}{0.0867} & 0.1416 & 0.1318 & 0.1009 & 0.1531 & 0.0883 &\color{green}{0.0865} &\textcolor{red}{0.0827}  \\ \cline{2-12} 
\multicolumn{1}{c|}{}                          & PSNR$\uparrow $  & 28.68  & 28.85  & 26.28  & 28.82  & 24.97  & \textbf{27.23}  & 25.48  & 27.11  & 27.02  & 27.14   \\ \cline{2-12} 
\multicolumn{1}{c|}{}                          & SSIM$\uparrow $  & 0.7883 & 0.7885 & 0.6980 & 0.7918 & 0.6908 & \textbf{0.7432} & 0.6889 & 0.7316 & 0.7257 & 0.7341  \\ \midrule
\multicolumn{1}{c|}{\multirow{3}{*}{BSD100}}   & LPIPS$\downarrow $ & 0.1517 & 0.1536 & \color{red}{0.0834} & 0.1497 & 0.1831 & 0.1041 & 0.1392 &\color{green}{0.0935} &\color{blue}{0.0953} & \color{red}{0.0834}  \\ \cline{2-12} 
\multicolumn{1}{c|}{}                          & PSNR$\uparrow $  & 27.73  & 27.74  & 25.29  & 27.86  & 24.65  & 25.95  & 25.21  & 25.74  & 25.87  &\textbf{25.96}   \\ \cline{2-12} 
\multicolumn{1}{c|}{}                          & SSIM$\uparrow $  & 0.7425 & 0.7430 & 0.6495 & 0.7466 & 0.6573 &\textbf{0.6833} & 0.6498 & 0.6681 & 0.6706 & 0.6743  \\ \midrule
\multicolumn{1}{c|}{\multirow{3}{*}{Urban100}} & LPIPS$\downarrow $ & 0.1251 & 0.1220 &\color{green}{0.0944} & 0.1185 & 0.1279 & 0.1077 & 0.1993 &\color{blue}{0.0997} &\color{blue}{0.0997} &\color{red}{0.0934}  \\ \cline{2-12} 
\multicolumn{1}{c|}{}                          & PSNR$\uparrow $  & 26.65  & 26.75  & 24.35  & 27.08  & 23.65  & 25.34  & 22.49  & 25.45  & 25.59  &\textbf{25.86}   \\ \cline{2-12} 
\multicolumn{1}{c|}{}                          & SSIM$\uparrow $  & 0.8036 & 0.8066 & 0.7327 & 0.8165 & 0.7312 & 0.7661 & 0.6336 & 0.7649 & 0.7681 &\textbf{0.7796}  \\ \midrule
\multicolumn{1}{c|}{\multirow{3}{*}{Manga109}} & LPIPS$\downarrow $ & 0.0628 & 0.0544 & 0.0420 & 0.0592 & 0.0660 & 0.0473 & 0.1427 &\color{blue}{0.0409}   &\color{green}{0.0396} &\color{red}{0.0374}  \\ \cline{2-12} 
\multicolumn{1}{c|}{}                          & PSNR$\uparrow $  & 31.06  & 31.20  & 28.48  & 31.67  & 27.14  & 28.67  & 24.69  &29.07 & 29.39  &\textbf{29.60}  \\ \cline{2-12} 
\multicolumn{1}{c|}{}                          & SSIM$\uparrow $  & 0.9160 & 0.9170 & 0.8595 & 0.9227 & 0.8244 & 0.8851 & 0.7568 & 0.8791 & 0.8816 &\textbf{0.8874}  \\ 
\midrule
\bottomrule
\end{tabular}}
\end{center}
\vspace{-18pt}
\end{table*}

\begin{table}[!htbp]
  \begin{center}
  \vspace{-14pt}
    \caption{Results of ablation study for different conditional inputs on BSD100. (4 $\times$ SR)}
  \label{tab:different_conditional_inputs}
   \resizebox{0.9\hsize}{!}{
\begin{tabular}{c|c|c|c|c|c}
\toprule
\hline
\multirow{2}{*}{Method} & \multirow{2}{*}{LR}   & \multirow{2}{*}{SR-RRDB} &  \multicolumn{3}{c}{\textbf{ours}} \\ \cline{4-6}
&&& \textbf{EDSR+} & \textbf{RCAN+} & \textbf{SwinIR+}\\\midrule
LPIPS$\downarrow$  & 0.1412 & 0.1096  & 0.0935  & 0.0953  &\textbf{ 0.0834}    \\ \hline
PSNR$\uparrow$   & 24.35  & 25.01   & 25.74   & 25.87   &\textbf{ 25.96 }    \\ \hline
SSIM$\uparrow$    & 0.6402 & 0.6589  & 0.6681  & 0.6706  &\textbf{ 0.6743 }   \\ \hline
\bottomrule
\end{tabular}}
  \end{center}
  \vspace{-24pt}
\end{table}

\vspace{-10pt}
\subsection{Ablation Study}
\label{ab:different_conditional_inputs}
\vspace{-4pt}

\begin{figure}[!ht]
\centering	
    \includegraphics [width=1\linewidth]{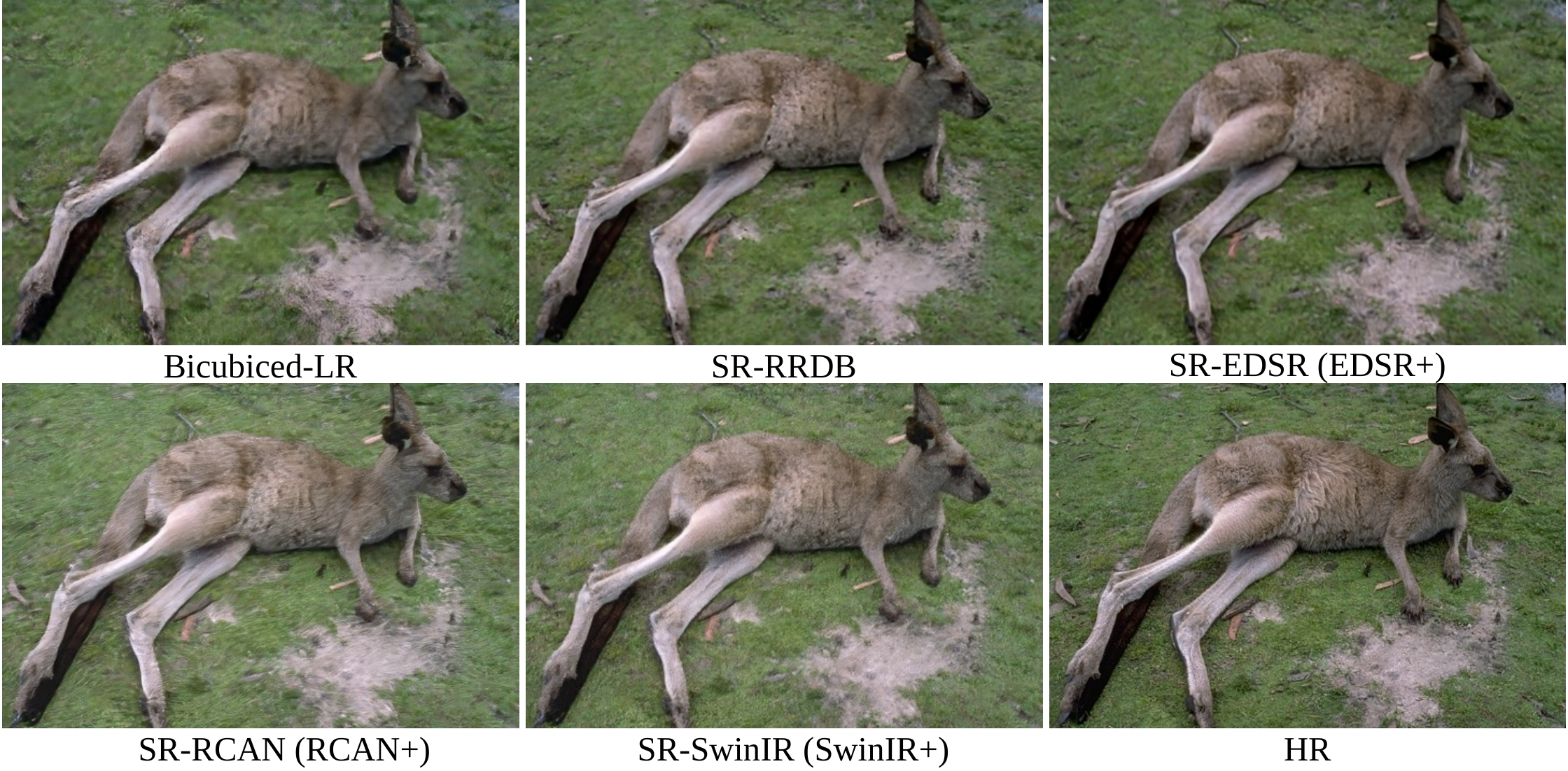}
    \vspace{-20pt}
    \caption{Visual results on '69020' from BSD100 under different conditional inputs ($\times$4 scale. Best view with zoomed-in.)}
    \label{fig:different_conditional_inputs}
\vspace{-16pt}
\end{figure}

\textbf{Different conditional inputs.} Here, we conduct experiments to verify how different conditional inputs influence performance. We adopt LR, SRs generated by EDSR, RCAN, SwinIR, and the RRDB trained in~\cite{li2022srdiff} as the conditional inputs to perform experiments, respectively. As shown in Tab.\ref{tab:different_conditional_inputs} and Fig.\ref{fig:different_conditional_inputs}, without any pre-processing, the result under LR conditional performs worst in both quantitative and qualitative. After being pre-trained by RRDB, EDSR, RCAN, SwinIR, the conditional inputs can restore more details, which pushes our cDPMSR model gets better performance.

\vspace{-10pt}
\section{Conclusion}
Our work revisits DPM in SR and reveals taking a pre-super-resolved version for the given LR image as the conditional input can help to achieve a better high-resolution image. Based on this, we propose a simple but non-trivial DPM-based super-resolution post-process framework,\ie, cDPMSR. By taking a pre-super-resolved version of the given LR image and adapting the standard DPM to perform super-resolution, our cDPMSR improves both qualitative and quantitative results and can generate more  photo-realistic counterparts for the low-resolution images on benchmark datasets (Set5, Set14, Urban100, BSD100, Manga109). In the future, we will extend our cDPMSR to images with more complex degradation.

\clearpage

\vfill\pagebreak

\bibliographystyle{IEEEbib}
\bibliography{strings,refs}

\end{document}